\documentclass[runningheads]{llncs}
\usepackage[T1]{fontenc}
\usepackage{graphicx}
\usepackage{amsmath}
\usepackage{amssymb}
\usepackage{booktabs}
\usepackage{algorithm}
\usepackage{algpseudocode}
\usepackage{float}
\usepackage{multirow}
\newcommand{\ProcName}[1]{\textnormal{#1}}
\algrenewcommand{\algorithmiccomment}[1]{%
  \hspace{2.6em}\texttt{//}\hspace{0.3em}#1%
}
\usepackage{booktabs}
\usepackage{tabularx}
\usepackage{cite}
\begin{document}
\title{CW-Ghost: Search-Free Granularity Selection for
Helper-Thread Prefetching via Capacity Windows}
\titlerunning{CW-Ghost: Search-Free Granularity Selection}
\author{Ya Zhang\inst{1,2}\orcidID{0009-0005-6625-6214},
Tong Lei\inst{1,2}\orcidID{0009-0007-9748-5997} \\
Yao Chen\inst{1,2}\orcidID{0000-0002-8748-4508} \and
Yonggang Che\thanks{Corresponding author}\inst{1,2}\orcidID{0000-0001-6906-4940} \and
Chuanfu Xu\inst{1,2}\orcidID{0000-0002-4876-2368} \and
Haozhong Qiu\inst{1,2}\orcidID{0009-0009-0434-8075} \and \\
Yusong Tan\inst{1,2}\orcidID{0000-0003-1233-5679}}
\authorrunning{Y. Zhang et al.}
\institute{Laboratory of Digitizing Software for Frontier Equipment, National University of Defense Technology, Changsha 410073, China \and
National Key Laboratory of Parallel and Distributed Computing, College of Computer Science and Technology, National University of Defense Technology, Changsha 410073, China\\
\email{ygche@nudt.edu.cn}}
\maketitle             

\begin{abstract}
Helper-thread prefetching hides the latency of irregular memory accesses by
executing address dependency chains ahead of the main thread. However, its effectiveness depends on the range of future iterations covered by
the helper thread. A fixed coverage range cannot consistently accommodate different workloads and processors, whereas exhaustively evaluating candidate configurations incurs substantial configuration cost. This paper presents  CW-Ghost,
which uses a single offline profiling run to estimate the average demand
cache line fill volume generated per target iteration in a target region. CW-Ghost combines this estimate with a cache capacity budget to derive a Capacity Window, which determines the iteration granularity of each prefetch chunk. In addition, bounded chunk-level synchronization limits the
number of chunks by which the helper thread may run ahead of the main thread. Across 14 workload instances evaluated on Intel and AMD CPU platforms, CW-Ghost achieves geometric mean speedups of \(1.54\times\) and
\(1.33\times\), respectively, over the original programs. Compared with Ghost
Threading, it improves geometric mean performance by \(15.8\%\) and
\(10.8\%\), respectively, while achieving more than \(99\%\) of the
empirically optimal performance within the candidate set on both platforms. These results demonstrate that cache capacity constraints can effectively
guide the selection of granularity for helper-thread prefetching.

\end{abstract}

\section{Introduction}

Applications such as graph analysis, databases, and high performance computing commonly involve irregular memory accesses, where the target address typically depends on preceding index accesses or data-dependent address calculation chains~\cite{guo2025ghost,ainsworth2019software,beamer2015gap,bailey1991nas}. Such dependences delay the generation of future memory requests, making
it difficult for conventional hardware prefetchers to consistently hide
memory access latency~\cite{ainsworth2019software,yu2015imp,fu2024differential}. Existing research has addressed such memory accesses through dedicated hardware prefetching or data access acceleration mechanisms~\cite{yu2015imp,fu2024differential}, software and compiler
prefetching~\cite{jamilan2022apt,ainsworth2019software,zhang2024rpg2,callahan1991software,mowry1992design,fu2025magellan,luk1996compiler}, and helper-thread prefetching or
runahead execution~\cite{kim2004physical,luk2001tolerating,jung2006helper,kamruzzaman2011inter,hashemi2016continuous,naithani2020precise}. Ghost Threading~\cite{guo2025ghost} targets existing SMT processors by
extracting an address-computation p-slice for a target load and running
a helper thread on a sibling SMT context to prefetch data for the main
thread's future accesses. On real Intel systems, it achieved a geometric mean speedup of \(1.33\times\) compared to the single-threaded baseline. However, synchronization parameters, including the inter-thread distance, still require manual tuning through profiling~\cite{guo2025ghost}.

Even after offloading address computation to a helper thread, it remains necessary to determine how many future target iterations that thread should cover at a time. We refer to a consecutive group of future target iterations
processed by the helper thread at a time as a \emph{prefetch chunk}, and define its
granularity, denoted by \(B\), as the number of target iterations in the chunk. When \(B\) is too small, prefetches may not be issued sufficiently far in advance, while chunk boundary checks and synchronization account
for a larger fraction of the execution overhead. When \(B\) is too large,
the expanded prefetch footprint can cause premature prefetching, data being evicted before it is used, and increased contention for caches, memory bandwidth, and SMT-shared resources~\cite{guo2025ghost,ainsworth2019software}. Consequently, neither a single fixed granularity nor a configuration selected on a source platform is consistently effective. Evaluating all candidate
granularities for every workload--platform pair, however, incurs
substantial configuration cost.

This paper proposes CW-Ghost, which replaces candidate-by-candidate
performance search for prefetch-chunk granularity with a capacity-guided
window computation. CW-Ghost uses a single offline profiling run to
estimate the demand cache line fill volume generated per target iteration
within the target region. It combines this estimate with a capacity budget
for the target cache to compute a Capacity Window, denoted by \(W_C\).
A prefetch chunk determined by \(W_C\) is defined as a
\emph{Capacity-Bounded Chunk} (CBC), whose granularity is set to $B_{\mathrm{CBC}} = W_C$. The optimized program partitions the target iteration space according to
this granularity and uses an independent CBC-level synchronization bound,
\(K\), to limit the chunk-level runahead of the helper thread. Thus,
\(W_C\) determines the amount of work covered by each prefetch chunk, whereas \(K\) controls the progress of the helper thread relative
to the main thread.

We evaluated CW-Ghost using 14 workload instances on two x86 SMT hardware
platforms equipped with an Intel Xeon Gold 6258R and an AMD EPYC 7H12.
Compared with the original single-threaded programs, CW-Ghost achieves
geometric mean speedups of \(1.54\times\) and \(1.33\times\) on the two platforms, respectively. Compared with Ghost Threading, it
improves geometric mean performance by \(15.8\%\) and \(10.8\%\),
respectively. Oracle-Chunk exhaustively evaluates a predefined candidate
set and selects the best observed granularity for each
workload--platform pair. CW-Ghost achieves more than \(99\%\) of
Oracle-Chunk's geometric mean performance on both platforms, while also
reducing the process-level retired instruction count and the main
thread's L1D-miss and LLC-miss MPKI.

The main contributions of this paper are as follows:

1. We characterize the performance impact of helper thread
prefetch-chunk granularity. Granularity sensitivity and
cross-processor configuration migration experiments show that
near-optimal granularity is both workload- and processor-dependent
and therefore cannot be reliably replaced by a single fixed value
or a configuration selected on another processor.

2. We propose the Capacity Window and CW-Ghost. CW-Ghost computes
the CBC granularity from the demand cache line fill volume per target
iteration obtained through offline profiling and a cache capacity
budget. It further uses bounded chunk-level synchronization to
control the progress of the helper thread.

3. We evaluate CW-Ghost on two x86 SMT hardware platforms based on
Intel Cascade Lake and AMD Zen~2, which differ in microarchitecture,
cache organization, and PMU event semantics. The results show that
CW-Ghost improves the end-to-end performance of Ghost Threading and
achieves overall performance close to the best observed
configuration in the candidate set.

\section{Background and Related Work}

\subsection{Challenges in Prefetching Indirect Memory Accesses}

Data-intensive applications, such as graph analytics and databases,
commonly contain a large number of indirect memory accesses. A
representative access pattern is \(A[B[i]]\), in which the processor must
first obtain the index \(B[i]\) before it can generate the address of the
corresponding access to \(A\)~\cite{guo2025ghost,ainsworth2019software,beamer2015gap, fu2025magellan,Guo2018DSAP,Lu2021GraphPEG}. Because the target address depends on runtime data, such accesses delay
the generation of subsequent memory requests, making it difficult for
conventional hardware prefetchers to request the target data sufficiently
early before the corresponding demand accesses~\cite{ainsworth2019software,yu2015imp,fu2024differential}.

Prefetching effectiveness depends on request accuracy and timeliness, as
well as the overhead of address generation and request
injection~\cite{guo2025ghost,jamilan2022apt,ainsworth2019software}. Prefetches issued too late cannot
sufficiently hide memory access latency, whereas those issued too early
may cause the prefetched data to be evicted before use and increase cache
and memory bandwidth pressure. Existing hardware mechanisms extend
address prediction capability by exploiting local delta~\cite{navarro2022berti},
temporal correlations~\cite{ainsworth2024triangel}, and online learning~\cite{bera2021pythia}.
Other designs employ specialized structures to capture indirect-access
relationships or accelerate address generation~\cite{yu2015imp,fu2024differential}.
Software prefetching~\cite{ainsworth2019software,callahan1991software,mowry1992design} and execution-based prefetching~\cite{luk2001tolerating,hashemi2016continuous}, on the other hand, explicitly generate
future memory requests by executing address computation chains ahead of the demand accesses.

\subsection{Software Prefetching}

Software prefetching relies on programmers or compilers to insert
explicit prefetch instructions into a program and compute the addresses
of data that will be accessed in the future~\cite{ainsworth2019software,callahan1991software,mowry1992design}.
Early studies established fundamental compiler techniques for
automatically inserting and scheduling prefetches~\cite{callahan1991software,mowry1992design},
and subsequent work extended these techniques to recursive and
pointer-based data structures~\cite{luk1996compiler}. For indirect memory accesses,
Ainsworth and Jones extract the address computation chain of a target
load and replicate the relevant instructions to generate prefetch
requests for subsequent loop iterations~\cite{ainsworth2019software}. Magellan further
exploits loop structures and inter-loop dependences to identify and
schedule both intra-loop and cross-loop prefetches for indirect memory
accesses, thereby expanding the range of address dependences that can be
handled by software prefetching~\cite{fu2025magellan}.

The effectiveness of software prefetching is generally sensitive to the
insertion point and lead distance of prefetch instructions, and optimal
configurations may vary with the program's dynamic behavior and the
target microarchitecture. APT-GET uses dynamic runtime profiling
to select prefetch injection sites and distances, thereby improving
prefetch timeliness~\cite{jamilan2022apt}. RPG\textsuperscript{2} profiles a running
program, injects prefetches, and dynamically adjusts the prefetch
distance to reduce the performance loss caused by unfavorable
configurations~\cite{zhang2024rpg2}. Because these approaches still execute
address computation and prefetch instructions on the main thread, their
performance benefits must be balanced against the additional overhead of dynamic instructions and execution resources.

\subsection{Helper-Thread Prefetching}

Helper-thread prefetching executes a compact pre-execution slice, or
\emph{p-slice}, in a separate execution context. The p-slice consists of
a target long-latency load and the instructions required to compute its
address, allowing the helper thread to generate future memory requests
for the main thread in advance~\cite{luk2001tolerating}. Since a p-slice
is typically shorter than the corresponding execution path in the
original program, the helper thread can run ahead of the main thread and
increase memory-level parallelism. However, insufficient runahead cannot
effectively hide memory-access latency, whereas excessive runahead may
cause eviction before use, cache pollution, and contention for shared
execution resources. Prior studies have investigated helper-thread prefetching on real SMT
systems~\cite{kim2004physical}, compiler-assisted slice extraction~\cite{jung2006helper},
and helper threads that migrate across processor cores~\cite{kamruzzaman2011inter}.
Related runahead execution techniques speculatively expose subsequent
memory accesses within the processor rather than using a separate
software thread context~\cite{hashemi2016continuous,naithani2020precise}.

Ghost Threading~\cite{guo2025ghost} implements software helper-thread
prefetching on existing SMT processors. It uses profiling to select a
target load, extracts the corresponding p-slice, and executes a ghost
thread on a sibling SMT context. However, its synchronization and
inter-thread distance parameters still require manual profile-guided
tuning~\cite{guo2025ghost}. Building on this helper execution framework,
CW-Ghost focuses on the low-cost selection of prefetch-chunk granularity.

\section{CW-Ghost: Capacity-Guided Granularity Selection and
Bounded Execution}
\label{sec:cwghost}

This section first characterizes the workload and processor dependence
of prefetch-chunk granularity, and then presents the Capacity Window
computation, CBC-based program transformation, and bounded chunk-level
execution mechanisms of CW-Ghost.

\subsection{Problem Characterization and Design Goals}
\label{sec:characterization}

We refer to a consecutive group of future target iterations processed by
the helper thread as a \emph{prefetch chunk}. A prefetch chunk is a general helper execution unit with no predefined cache-capacity constraint; its granularity, denoted by \(B\), is the number of target iterations it contains. To determine the feasibility of uniform granularity or cross-processor reuse configurations, this section examines the workload dependency, processor dependency, and configuration cost of \(B\), while keeping other helper execution mechanisms unchanged. Here, \(B\) represents a general candidate granularity; the CBC granularity determined by the Capacity Window will be defined in Section~\ref{sec:cwghost-overview}.

\subsubsection{Workload Dependence}

\begin{figure}[t]
   \centering
   \includegraphics[width=\linewidth, trim=5 8 5 7, clip]{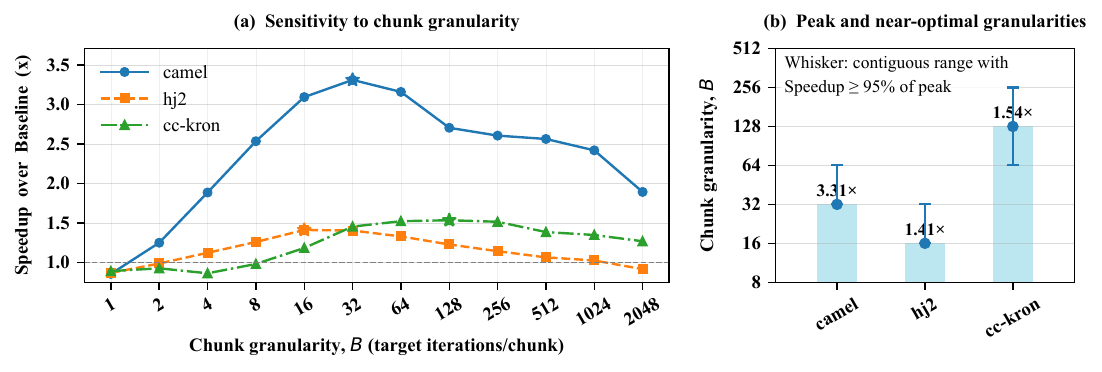}
   \caption{Workload dependence of prefetch-chunk granularity.
    (a) Speedup of \texttt{camel}, \texttt{hj2}, and
    \texttt{cc-kron} over Baseline on the Intel platform under
    different \(B \in \mathcal{B}\). Stars mark the peak configurations
    in the candidate set, and the horizontal dashed line denotes
    Baseline. (b) Peak granularities and near-optimal ranges. Circles
    mark the peak granularities, the annotations report the corresponding
    peak speedups, and each whisker indicates the largest contiguous
    candidate interval that contains the peak and achieves at least
    \(95\%\) of the peak performance.}
    \label{fig:chunk-sensitivity} 
 \end{figure}

To examine the effect of chunk granularity on performance, we fix the
target region, target load, address computation slice, chunk-level
synchronization mechanism, maximum runahead bound, and compilation
configuration on the Intel platform. We vary only \(B\) over the
following set of 12 power of two candidates, i.e.,
$\mathcal{B}=\{2^k \mid 0 \leq k \leq 11\}
=\{1,2,\ldots,2048\}$. All other experimental settings are identical to those used in the main
experiments described in Section~\ref{sec:experimental-setup}.
Figure~\ref{fig:chunk-sensitivity} shows that the performance of
\texttt{camel}, \texttt{hj2}, and \texttt{cc-kron} varies
non-monotonically with \(B\). Moreover, neither the peak granularity nor
the near-optimal range whose performance is at least \(95\%\) of the
corresponding peak is consistent across the three workloads. Therefore, a single fixed granularity cannot be reliably applied to different workloads.

\subsubsection{Processor Dependence}

To determine whether a chunk-granularity configuration can be reused
across processors, we evaluate the same candidate set on the Intel and
AMD platforms while holding the workload, input, target load,
address computation slice, synchronization mechanism, and compiler
optimizations constant. Let \(S_p(B)\) denote the speedup achieved on
platform \(p\) using granularity \(B\), relative to the original
single-threaded program on the same platform. The locally empirically optimal granularity for platform p is defined as:
\begin{equation}
B_p^*
=
\underset{B \in \mathcal{B}}{\arg\max}\; S_p(B).
\label{eq:local-optimal-granularity}
\end{equation}
The normalized transfer performance from a source platform \(p\) to a
target platform \(q\) is defined as:
\begin{equation}
T_{p \rightarrow q}
=
\frac{S_q\!\left(B_p^*\right)}
     {S_q\!\left(B_q^*\right)}.
\label{eq:normalized-transfer-performance}
\end{equation}
Here, \(T_{p \rightarrow q}=1\) indicates that the granularity selected
on the source platform achieves the target platform's local peak
performance, whereas \(T_{p \rightarrow q}<1\) indicates a transfer
loss. Both the numerator and denominator are measured on the target
platform relative to the same Baseline. The metric therefore reflects
the quality of the transferred granularity rather than differences in
absolute execution time between the two platforms.

\begin{figure}[t]
    \centering
    \includegraphics[width=\linewidth, trim=10 20 25 5, clip]
    {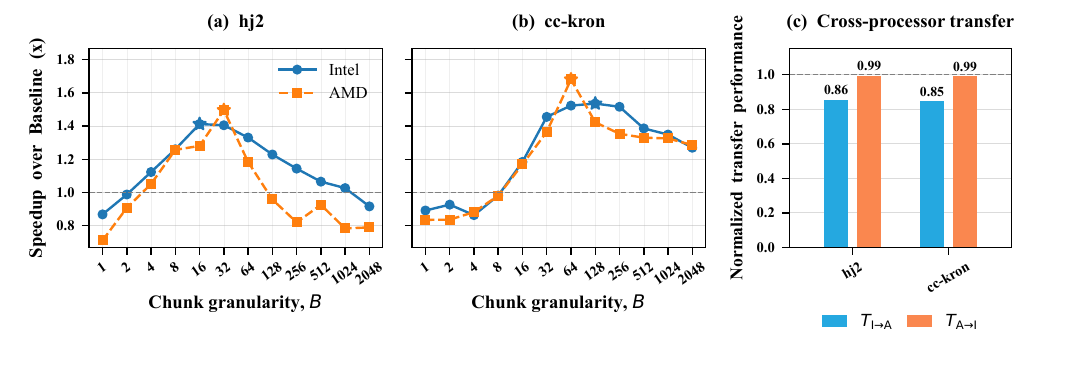}
    \caption{Cross-processor transferability of prefetch-chunk
    granularity. (a)--(b) Granularity sensitivity of \texttt{hj2} and
    \texttt{cc-kron} on the Intel and AMD platforms. Stars indicate
    the platform-local empirically optimal granularities within the
    predefined candidate set, and the horizontal dashed line denotes
    Baseline. (c) Bidirectional normalized transfer performance.
    \(T_{\mathrm{I}\rightarrow\mathrm{A}}\) and
    \(T_{\mathrm{A}\rightarrow\mathrm{I}}\) denote the normalized
    performance obtained by applying the Intel- and AMD-local optimal
    granularities, respectively, to the other platform.}
    \label{fig:cross-processor-transfer}
\end{figure}

As shown in Figure~\ref{fig:cross-processor-transfer}, the peak
granularity of the same workload can vary across processors. For
\texttt{hj2}, the platform-local empirically optimal granularities are
16 and 32 on the Intel and AMD platforms, respectively; for
\texttt{cc-kron}, they are 128 and 64, respectively. When the
Intel-local optimum is transferred to AMD, \texttt{hj2} and
\texttt{cc-kron} achieve 0.86 and 0.85 of the corresponding AMD-local
optimal performance. In the reverse direction, both workloads retain
approximately 0.99 of the Intel-local optimal performance. Transfer
quality depends not only on the locations of the peaks on the two
platforms, but also on whether the source-platform granularity falls
within the near-optimal range of the target platform. Consequently, a
granularity that is locally optimal on one processor is not necessarily
near-optimal on another.

\subsubsection{Configuration Cost}

Oracle-Chunk evaluates every granularity in \(\mathcal{B}\) and selects
the configuration with the highest measured performance for each
workload--platform pair. Although this procedure provides an empirical
performance upper bound within the candidate set, it requires all
candidates to be executed for every target region and platform. Its cost
further increases with the numbers of processors, inputs, and target
regions, making it unsuitable as a practical configuration procedure.

These observations lead to three design goals. First, granularity
selection should reflect the dynamic behavior of the current
workload--platform pair rather than rely on a single fixed value or a
configuration transferred from another platform. Second, the
configuration procedure should avoid evaluating candidate granularities
one by one. Third, profiling logic should not be included in the final
optimized program or affect its performance measurement.

\subsection{CW-Ghost Overview}
\label{sec:cwghost-overview}

\begin{figure}[t]
    \centering
    \includegraphics[width=\linewidth, trim=9 3 10 3, clip]
    {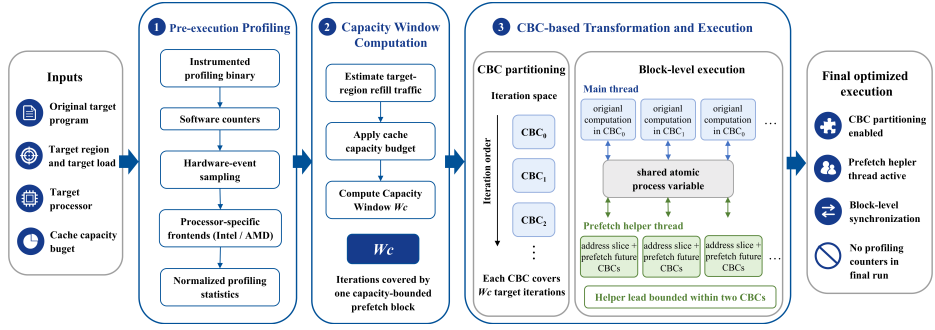}
    \caption{End-to-end workflow of CW-Ghost. Offline profiling produces
    a normalized profiling record, from which the window computation
    stage derives \(W_C\) and sets
    \(B_{\mathrm{CBC}} = W_C\). The final program partitions the
    target iteration space into CBCs using this granularity and limits
    the helper thread's chunk-level runahead through the synchronization
    bound \(K\).}
    \label{fig:cw-ghost-overview}
\end{figure}

Figure~\ref{fig:cw-ghost-overview} illustrates the three stages of
CW-Ghost: offline profiling, Capacity Window computation, and CBC-based
execution. The method takes as input a predetermined target region,
target load, and corresponding address computation slice. Target region
identification, target load selection, and generic slice extraction are beyond the scope of the automation provided by this
work.

During the offline profiling stage, CW-Ghost combines software iteration
counting with platform-specific PMU sampling to obtain the dynamic number
of target iterations, the number of cache event samples, and the sampling
period for the target region. These statistics are organized into a
unified normalized profiling record. Platform-specific differences are
confined to the event collection and report parsing front ends. All
subsequent stages use the same window computation logic, without
evaluating candidate chunk granularities one by one.

During the Capacity Window computation stage, CW-Ghost combines the
normalized profiling record with the cache configuration and capacity
budget of the target processor to compute the Capacity Window,
\(W_C\). We define a prefetch chunk whose granularity is determined by
\(W_C\) as a \emph{Capacity-Bounded Chunk} (CBC), and set
$B_{\mathrm{CBC}} = W_C$.
Here, \(B\) denotes the general candidate prefetch-chunk granularity
introduced in Section~\ref{sec:characterization}, whereas
\(B_{\mathrm{CBC}}\) denotes the CBC granularity computed by CW-Ghost.

During the CBC-based execution stage, the final optimized program
partitions the target iteration space according to
\(B_{\mathrm{CBC}}\) and runs the prefetch helper thread on a sibling
SMT context of the same physical core. The value
\(B_{\mathrm{CBC}}\) determines the number of target iterations covered
by each CBC, while an independent CBC-level synchronization bound
\(K\) limits the chunk-level runahead of the helper thread relative to
the main thread.

\subsection{Profiling and Capacity Window Computation}
\label{sec:capacity-window-computation}

The offline profiling stage of CW-Ghost collects only the statistics
required to compute the Capacity Window. For a target region \(R_t\),
profiling obtains the number of hardware event samples attributed to
the region \(S_t\), the sampling period \(I\), and the dynamic number
of target iterations \(N_t\). These statistics form the following
normalized profiling record:
\begin{equation}
\Pi_t = \left\langle S_t, I, N_t \right\rangle.
\label{eq:normalized-profile-record}
\end{equation}
Here, \(N_t\) denotes the dynamic number of target iterations covered by
the final CBC-based transformation and does not necessarily equal the
static trip count of an enclosing loop.

\subsubsection{Normalized Profile}

Hardware event sampling captures the distribution of cache events
across the program address space, but does not directly provide the
dynamic target iteration count \(N_t\). The profiling version therefore
uses conditionally compiled software counters to record \(N_t\). For a
region whose loop extent is directly available, the counter is
incremented in bulk by the actual loop extent. For a region whose
iteration count depends on the input or control flow, the actual number of target iterations executed is recorded. These software counters
are excluded from the final optimized program. Hardware samples are
attributed to the target region \(R_t\) using instruction addresses and
debug information.

\subsubsection{Platform-Specific Collection}

The Intel and AMD platforms use different front ends for hardware event
collection and report parsing. On Intel, CW-Ghost uses PEBS to collect
samples of retired demand loads that miss in the L1D cache. On AMD, it
uses EBS to collect demand data cache refill samples. The detailed event
definitions follow the corresponding vendor
documentation~\cite{amd_uprof_2024,intel_sdm}, while the counting scopes are
described in Section~\ref{sec:experimental-setup}. Although these
events have different semantics, both are used in the window model to approximate the L1D refill behavior caused by demand accesses in the target region. The model runs independently on each platform, and no absolute cross-processor comparisons are made between the number of raw samples and the number of reconstructed events.

\subsubsection{Capacity Window Model}

For sampling at fixed intervals, $S_t \cdot I$ estimates the total number of monitored events in the target region.
Let \(L\) denote the cache line size. The average demand cache line fill
volume per target iteration is then estimated as:
\begin{equation}
\widehat{F}_t
=
\frac{S_t \times I \times L}{N_t}.
\label{eq:fill-volume-per-iteration}
\end{equation}
The unit of \(\widehat{F}_t\) is bytes per target iteration.

CW-Ghost uses the physical L1D capacity, \(C_{\mathrm{L1D}}\), as the
modeling capacity because the sampled events approximate L1D
misses/refills and the main and helper threads running on the same
physical core share this cache level. Let \(\alpha \in (0,1]\) be the global capacity budget ratio. Under the condition that \(\widehat{F}_t\) approximates the refill volume for different target iterations within a chunk and is accumulated over iterations, the number of non-discretized target iterations that the capacity budget can cover is:
\begin{equation}
W_{\mathrm{raw}}
=
\frac{\alpha \times C_{\mathrm{L1D}}}{\widehat{F}_t}
=
\frac{
    \alpha \times C_{\mathrm{L1D}} \times N_t
}{
    S_t \times I \times L
}.
\label{eq:raw-capacity-window}
\end{equation}
We use the same value of \(\alpha\) for all workloads and processors, and its sensitivity is evaluated in Section~\ref{sec:capacity-window-effectiveness}.

\subsubsection{Validity and Discretization}
The model proceeds only when
$S_t > 0$, $I > 0$, $N_t > 0$,
$C_{\mathrm{L1D}} > 0$, $L > 0$.
If \(\widehat{F}_t \leq 0\) or \(W_{\mathrm{raw}} < 1\), CW-Ghost does
not generate a CBC configuration for the target region. Otherwise, the
valid result is rounded down to the largest power of two no greater than
\(W_{\mathrm{raw}}\):
\begin{equation}
W_C
=
2^{\left\lfloor \log_2 W_{\mathrm{raw}} \right\rfloor}.
\label{eq:capacity-window-discretization}
\end{equation}
This rounding ensures that the resulting \(W_C\) does not exceed
the upper bound given by the capacity model, while keeping chunk
boundaries and granularity representation regular. The final CBC
granularity is $B_{\mathrm{CBC}} = W_C$.

\subsubsection{Model Scope}

The quantity \(\widehat{F}_t\) is a cumulative fill volume proxy derived
from sampled events. It is not equivalent to the set of unique cache
lines, a strict working set size, or the actual cache residency at any given moment. Repeated fills, inter-iteration reuse, and
cache replacement may all cause it to differ from the actual cache
occupancy. The Capacity Window is therefore a granularity selection rule based on capacity budgets, rather than an exact predictor of cache state
or of the granularity that maximizes end-to-end performance.

\subsection{CBC-Based Program Transformation and Bounded Execution}
\label{sec:cbc-transformation}

CW-Ghost sets \(B_{\mathrm{CBC}} = W_C\), partitions the target iteration
space accordingly, and uses CBC-level synchronization to control the
relative progress of the main thread and the prefetch helper thread.

\subsubsection{CBC Partitioning}

For an execution containing \(N_t\) target iterations, CW-Ghost
partitions the target iterations into consecutive Capacity-Bounded
Chunks and uses a consistent CBC index space for the main and helper
threads. The total number of CBCs is:
\begin{equation}
N_{\mathrm{CBC}}
=
\left\lceil
\frac{N_t}{B_{\mathrm{CBC}}}
\right\rceil
=
\left\lceil
\frac{N_t}{W_C}
\right\rceil.
\label{eq:number-of-cbcs}
\end{equation}
For \(0 \leq c < N_{\mathrm{CBC}}\), the \(c\)th CBC covers the following
target-iteration interval:
\begin{equation}
\mathrm{CBC}_c
=
\left[
c B_{\mathrm{CBC}},
\;
\min( (c+1)B_{\mathrm{CBC}},\, N_t )
\right).
\label{eq:cbc-interval}
\end{equation}
Except for the final CBC, each CBC contains exactly
\(B_{\mathrm{CBC}}\) target iterations. The main thread executes CBCs
in increasing index order while preserving the original iteration
order within each CBC. The transformation introduces only chunk boundary
computation and progress publication, without changing the original
computation semantics of the main thread.

\subsubsection{Helper Execution}

The helper thread uses the same CBC index space and executes the
target address computation slice \(A_t\) ahead of the main thread for
future CBCs. The slice preserves the control flow, index computations,
and data dependences required to generate the target addresses, replaces the target load with a prefetch hint, and does not execute its downstream consumer semantics.

\subsubsection{Bounded Progress Control}

\begin{figure}[t]
    \centering
    \includegraphics[width=\linewidth, trim=25 15 25 15, clip]
    {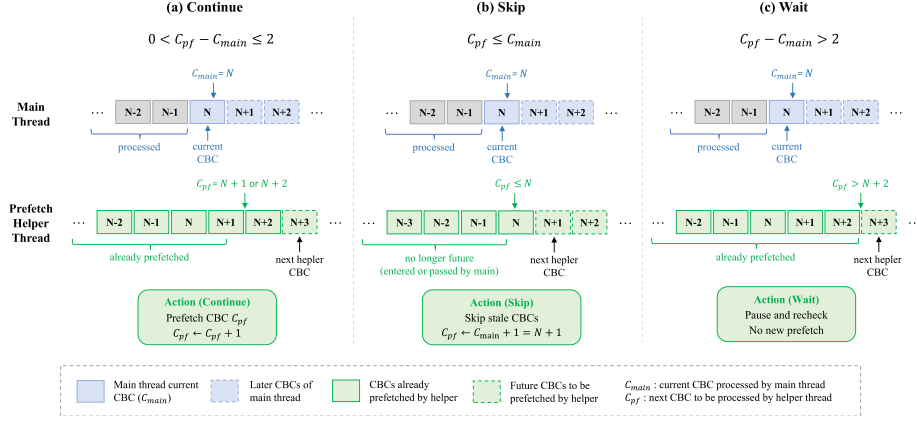}
    \caption{CBC-level bounded runahead control. The main thread
    publishes the index of its current CBC through
    \(C_{\mathrm{main}}\), while the helper thread maintains the next
    CBC to be processed in \(C_{\mathrm{pf}}\). Based on their relative
    positions, the helper performs Continue, Skip, or Wait.}
    \label{fig:bounded-cbc-runahead}
\end{figure}

While \(B_{\mathrm{CBC}}\) determines the number of target iterations
covered by each CBC, an independent CBC-level synchronization bound \(K\) restricts the range of future CBCs that the helper thread may
select. The two threads share an atomic variable \(C_{\mathrm{main}}\),
which denotes the CBC currently being executed by the main thread. The
main thread is the sole writer of \(C_{\mathrm{main}}\), while the helper
thread only reads it and locally maintains \(C_{\mathrm{pf}}\), the index
of the next CBC it intends to process. Valid CBC indices lie in
\([0,N_{\mathrm{CBC}}-1]\). Before execution begins, the main thread
initializes \(C_{\mathrm{main}}\) to 0. After completing the target
region, it updates \(C_{\mathrm{main}}\) to \(N_{\mathrm{CBC}}\), which
serves as the completion sentinel. The main thread publishes its progress
only when entering a new CBC. The helper thread reads the shared variable
when selecting the next CBC or waiting for the main thread to advance.

As shown in Figure~\ref{fig:bounded-cbc-runahead}, the helper thread skips CBCs where \(C_{\mathrm{pf}} \leq C_{\mathrm{main}}\), continues prefetching when \(0 < C_{\mathrm{pf}} - C_{\mathrm{main}} \leq K\), and waits for the main thread to advance when \(C_{\mathrm{pf}} > C_{\mathrm{main}}+K\). Algorithm~\ref{alg:bounded-cbc-execution} summarizes the complete process of CBC progress publication by the main thread and bounded prefetching by the helper thread. The two threads join only after the target region has completed.

The synchronization bound \(K\) limits the maximum chunk-level lead
when the helper selects new work relative to the main thread progress
most recently observed by the helper. Specifically, a newly selected
CBC satisfies \(0 < C_{\mathrm{pf}} - C_{\mathrm{main}} \leq K\).
This relation need not remain valid throughout the helper's execution
of the selected CBC. The main thread may continue to advance while the
helper executes its slice. Before selecting the next CBC, the helper
reloads the main thread progress and uses the \(Skip\) action to bypass CBCs
that the main thread has already entered. Consequently, \(K\) is neither
a strict iteration-level prefetch distance nor an upper bound on the
aggregate cache residency of prefetched but unconsumed data. It merely constrains the relative CBC-level progress when new helper work is selected.

\subsubsection{Correctness and Scope}

The CBC-based transformation requires \(A_t\) to commit no state modifications or external side effects visible to the application and to
introduce neither invalid memory accesses nor undefined data races. The
main and helper threads must also establish a consistent and monotonic
index mapping for the target iterations. Shared progress is communicated
through an atomic variable using release semantics for main thread stores. However, this communication does not replace the data safety conditions required by
\(A_t\) itself. For target regions that do not satisfy the above conditions, CW-Ghost does not apply this transformation.

\begin{algorithm}[H]
\caption{Bounded CBC-Based Main/Helper Execution}
\label{alg:bounded-cbc-execution}
\begin{algorithmic}[1]

\Statex \textbf{Input:}
\Statex \hspace{\algorithmicindent}
        \makebox[1.3cm][l]{$N_t$}
        \texttt{//} number of target iterations in the current execution
\Statex \hspace{\algorithmicindent}
        \makebox[1.3cm][l]{$B_{\mathrm{CBC}}$}
        \texttt{//} CBC granularity, set to $W_C$
\Statex \hspace{\algorithmicindent}
        \makebox[1.3cm][l]{$A_t$}
        \texttt{//} pre-extracted target-address computation slice
\Statex \hspace{\algorithmicindent}
        \makebox[1.3cm][l]{$K$}
        \texttt{//} CBC-level synchronization bound, $K\geq1$

\Statex

\Statex \textbf{procedure}
        \ProcName{MainThread}$(N_t,B_{\mathrm{CBC}},A_t,K)$
\State $N_{\mathrm{CBC}}
       \gets\left\lceil N_t/B_{\mathrm{CBC}}\right\rceil$
\If{$N_{\mathrm{CBC}}=0$}
  \State \Return
\EndIf
\State $C_{\mathrm{main}}.\mathrm{store}
       (0,\mathrm{release})$
\State \textbf{launch}
       \ProcName{PrefetchHelper}
       $(N_{\mathrm{CBC}},B_{\mathrm{CBC}},A_t,K)$
\State \ProcName{ExecuteOriginal}$(\mathrm{CBC}_0)$
\For{$c\gets1$ \textbf{to} $N_{\mathrm{CBC}}-1$}
  \State $C_{\mathrm{main}}.\mathrm{store}
         (c,\mathrm{release})$
  \State \ProcName{ExecuteOriginal}$(\mathrm{CBC}_c)$
\EndFor
\State $C_{\mathrm{main}}.\mathrm{store}
       (N_{\mathrm{CBC}},\mathrm{release})$
       \Comment{completion sentinel}
\State \textbf{join} \ProcName{PrefetchHelper}

\Statex

\Statex \textbf{procedure}
        \ProcName{PrefetchHelper}
        $(N_{\mathrm{CBC}},B_{\mathrm{CBC}},A_t,K)$
\State $C_{\mathrm{pf}}\gets0$
\While{$C_{\mathrm{pf}}<N_{\mathrm{CBC}}$}
  \State $c\gets
         C_{\mathrm{main}}.\mathrm{load}(\mathrm{acquire})$

  \If{$c\geq N_{\mathrm{CBC}}$}
    \State \textbf{break}
  \EndIf

  \If{$C_{\mathrm{pf}}\leq c$}
    \State $C_{\mathrm{pf}}\gets c+1$ \Comment{Skip}
  \EndIf

  \If{$C_{\mathrm{pf}}\geq N_{\mathrm{CBC}}$}
    \State \textbf{break}
  \EndIf

  \If{$C_{\mathrm{pf}}>c+K$}
    \State \ProcName{Pause}() \Comment{Wait}
    \State \textbf{continue}
  \EndIf

  \State \ProcName{PrefetchCBC}
         $(\mathrm{CBC}_{C_{\mathrm{pf}}},A_t)$
         \Comment{Continue}
  \State $C_{\mathrm{pf}}\gets C_{\mathrm{pf}}+1$
\EndWhile

\end{algorithmic}
\end{algorithm}

\section{Experimental Evaluation}
\label{sec:evaluation}

\subsection{Experimental Setup}
\label{sec:experimental-setup}

\subsubsection{Platforms and Workloads}

Table~\ref{tab:platform-configuration} summarizes the two hardware platforms:
an Intel Xeon Gold 6258R based on the Cascade Lake microarchitecture and
an AMD EPYC 7H12 based on the Zen~2 microarchitecture. SMT is enabled on
both platforms. Each experiment uses one main thread and one prefetch
helper thread, which are pinned to sibling SMT contexts on the same
physical core. CW-Ghost performs offline profiling and computes the
Capacity Window \(W_C\) independently for each workload--platform pair.

Table~\ref{tab:workload-configuration} lists the evaluated programs, inputs, and
target regions. The evaluation suite includes graph workloads, an HPC
kernel, synthetic indirect memory access programs, and database hash
joins, comprising 14 workload instances in total. We refer to the combination of a kernel and a specific input as a workload. For
each workload, hardware sample attribution, dynamic target iteration
counting, and the CBC-based transformation refer to the same target
region. Unless otherwise stated, offline profiling and performance
evaluation use the same input.

\subsubsection{Configurations}
We evaluate the following five configurations:
\begin{itemize}
    \item \textbf{Baseline.}
    The single-threaded program without helper thread prefetching.

    \item \textbf{Ghost Threading.}
    Our port of Ghost Threading~\cite{guo2025ghost} to the evaluated hardware
    platforms. It adopts the helper thread execution framework and
    synchronization mechanism of the original design, with the
    adaptations required for each target platform. The relevant
    parameters follow the recommendations of the original work and the
    settings used in our port.

    \item \textbf{CW-Ghost.}
    CW-Ghost computes \(W_C\) through a single
    offline profiling run, sets
    \(B_{\mathrm{CBC}}=W_C\), and applies the CBC-based program
    transformation and bounded-execution mechanism described in
    Section~\ref{sec:cbc-transformation}.

    \item \textbf{Fixed-Chunk.}
    A fixed granularity shared across all workloads on the same platform, used to evaluate the effect of the Capacity Window relative to that fixed granularity.

    \item \textbf{Oracle-Chunk.}
    Selects the granularity with the lowest mean execution time from the predefined candidate set \(\mathcal{B}\) as the empirical performance upper
    bound within the candidate set. This configuration is used only to evaluate
    granularity-selection quality; it is not part of the practical
    configuration procedure of CW-Ghost.
\end{itemize}

\begin{table}[t]
\caption{Platform configuration.}
\label{tab:platform-configuration}
\centering
\small
\begin{tabularx}{\textwidth}{|l|X|X|}
\hline
\textbf{Component}
& \textbf{Intel platform}
& \textbf{AMD platform} \\
\hline
Processor
& Intel Xeon Gold 6258R (Cascade Lake)
& AMD EPYC 7H12 (Zen~2, Rome) \\
\hline
CPU topology
& $2$ sockets $\times$ $28$ cores/socket $\times$ $2$ threads/core
& $2$ sockets $\times$ $64$ cores/socket $\times$ $2$ threads/core \\
\hline
L1D cache
& 32~KiB/core
& 32~KiB/core \\
\hline
L2 cache
& 1~MiB/core
& 512~KiB/core \\
\hline
LLC organization
& 38.5~MiB/socket, shared
& 16~MiB/4-core CCX, shared \\
\hline
Main memory
& 376~GB DDR4
& 503~GB DDR4 \\
\hline
\end{tabularx}
\end{table}

\begin{table}[t]
\caption{Workload configuration.}
\label{tab:workload-configuration}
\centering
\small
\begin{tabularx}{\textwidth}{|p{1.8cm}|X|X|}
\hline
\textbf{Kernel}
& \textbf{Input(s)}
& \textbf{Target region} \\
\hline
cc
& kron, twitter, urand, road, web
& Afforest sampled-neighbor linking \\
\hline
sssp
& kron, twitter, urand, web
& $\Delta$-stepping edge relaxation \\
\hline
camel
& $2^{25}$ elements
& Main pointer-indirect traversal \\
\hline
kangaroo
& $2^{25}$ keys, three indirect arrays
& rank two-level hash-indirect update \\
\hline
is
& Class B, $2^{25}$ keys
& rank histogram update \\
\hline
hj2
& $|R|=|S|=12.8$ million tuples
& NPO hash-table probe \\
\hline
hj8
& $|R|=|S|=12.8$ million tuples
& NPO hash-table probe \\
\hline
\end{tabularx}
\end{table}

Fixed-Chunk and Oracle-Chunk reuse the target regions, address computation slices,
program transformation, and CBC-level synchronization mechanism of
CW-Ghost, and differ only in granularity selection. Except for the differences in the mechanisms described above, all
configurations on the same platform use identical inputs, main thread
computation, compiler options, thread and NUMA placement,
hardware prefetcher configuration, and timing scope.

\subsubsection{Parameters and Measurement}

Unless otherwise stated, the capacity budget ratio is set to
\(\alpha=0.5\), and the CBC-level synchronization bound is set to
\(K=2\). Both parameters remain fixed across all workloads and both
platforms, without per-workload or per-platform tuning. Their
sensitivity is evaluated in
Section~\ref{sec:capacity-window-effectiveness}.

We use kernel-level end-to-end execution time as the primary performance
metric. The timed region excludes input loading, test data generation,
and one-time initialization, but includes the complete computation
kernel and all overhead associated with helper thread creation, address
computation, prefetching, CBC-level synchronization, waiting, thread
termination, and thread joining. Each configuration is executed three
times, with each run launched as a separate process. Speedup over
Baseline is computed using the mean execution times, and speedups across
workloads are summarized using the geometric mean.

\subsubsection{PMU Metrics}

Performance timing and PMU metrics are collected in separate runs to
prevent hardware event collection from affecting the timing results.
Dynamic instruction counts are collected at process scope and include
all retired instructions from both the main and helper threads. For
cache MPKI, both the cache event counts and the retired instruction
denominators are collected at main thread scope. Because the Intel and
AMD platforms use PMU events with different semantics, these metrics are
normalized to Baseline only within the same platform; absolute event
counts are not compared across processors.

\subsection{End-to-End Performance}
\label{sec:end-to-end-performance}

\begin{figure}[t]
  \centering
  \begin{minipage}[b]{0.96\linewidth}
    \centering
    \includegraphics[width=\linewidth, trim=5 7 5 5, clip]{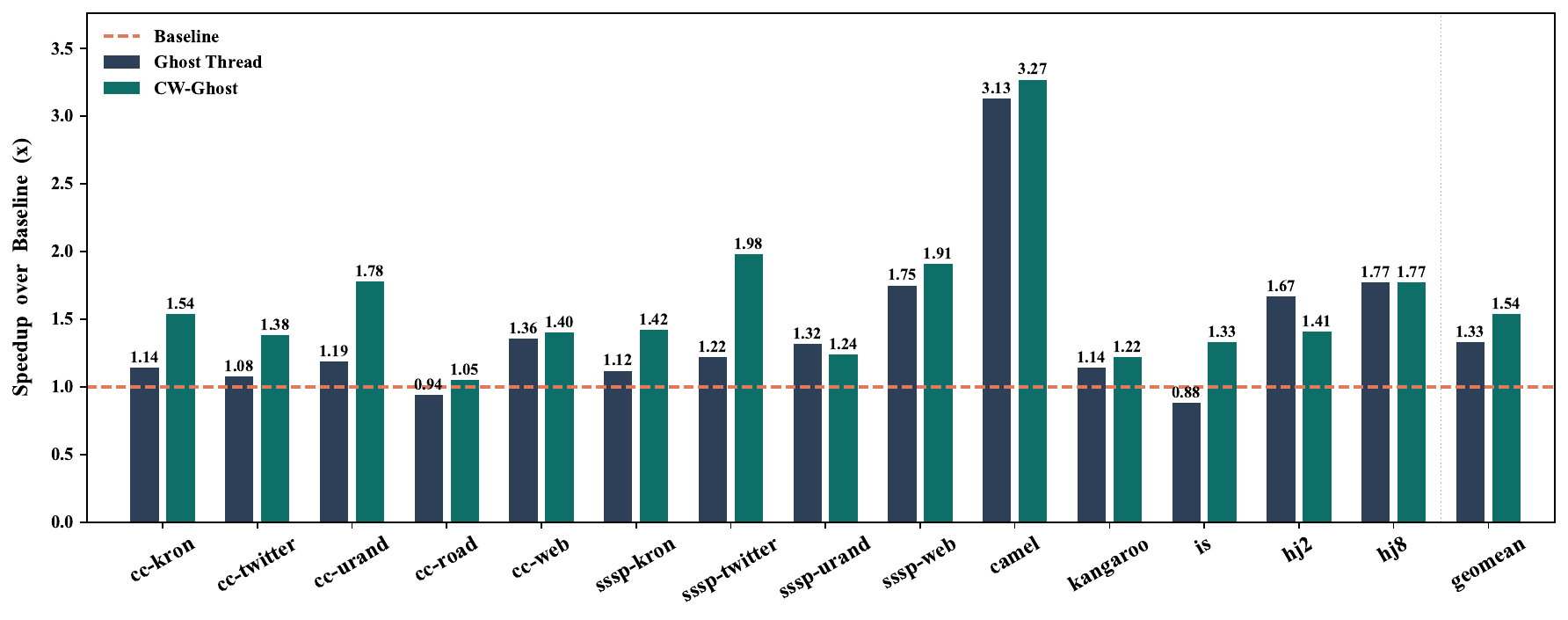}
    \textnormal{(a) Intel Xeon Gold 6258R}
  \end{minipage}

  \begin{minipage}[b]{0.96\linewidth}
    \centering
    \includegraphics[width=\linewidth, trim=5 7 5 5, clip]{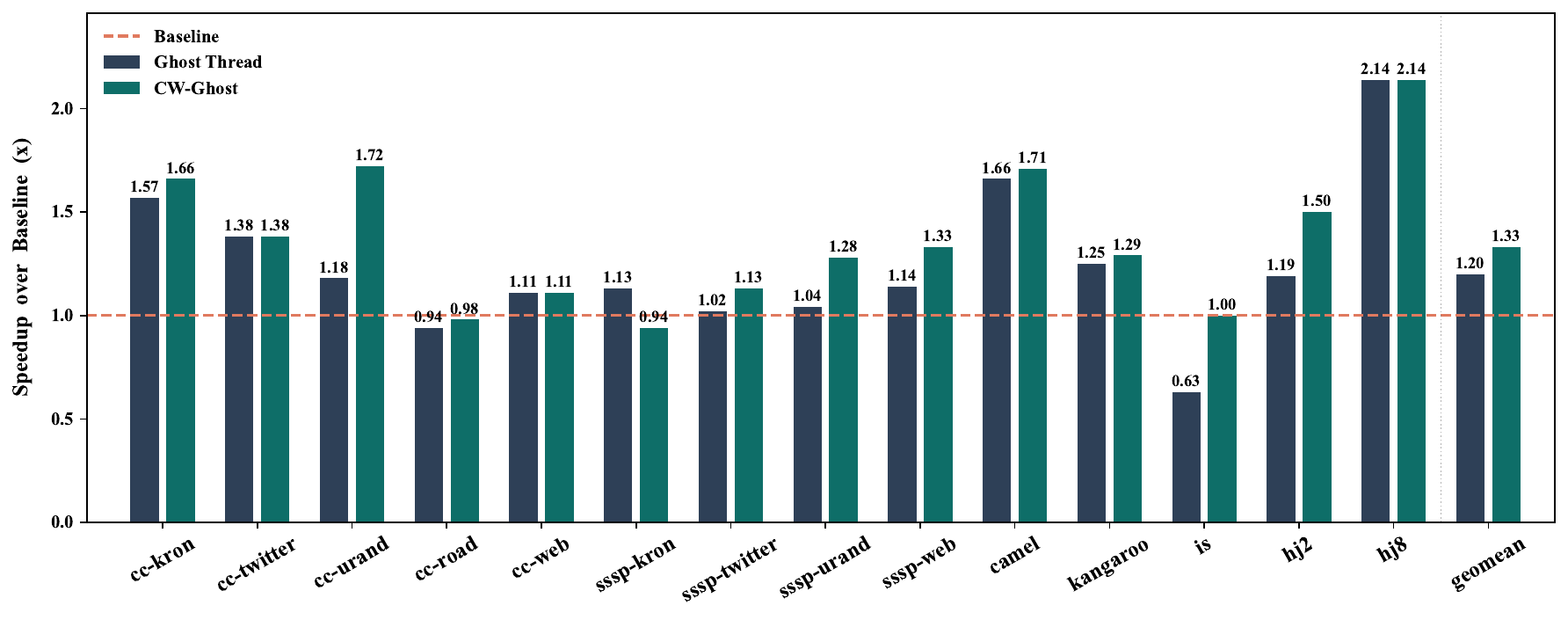}
    \textnormal{(b) AMD EPYC 7H12}
  \end{minipage}

  \caption{Kernel-level end-to-end speedup of Ghost Threading and CW-Ghost over
    Baseline on (a) the Intel Xeon Gold 6258R and (b) the AMD EPYC
    7H12. The geometric mean is calculated across all evaluated
    workloads. Higher is better.}
  \label{fig:end-to-end-speedup}
\end{figure}

Figure~\ref{fig:end-to-end-speedup} reports the kernel-level end-to-end
speedup of Ghost Threading and CW-Ghost over Baseline. On the Intel and AMD
platforms, CW-Ghost achieves geometric mean speedups of
\(1.54\times\) and \(1.33\times\), respectively, compared with
\(1.33\times\) and \(1.20\times\) for Ghost Threading. Thus, relative to Ghost Threading,
CW-Ghost improves geometric mean performance by \(15.8\%\) on Intel
and \(10.8\%\) on AMD.

Across the 14 evaluated workloads, CW-Ghost matches or outperforms Ghost Threading
on 12 workloads on the Intel platform and 13 workloads on the AMD
platform. This broad distribution indicates that  the performance gains covered the majority of workloads. Nevertheless,
individual workloads still exhibit limited gains or slight performance
degradation.

\subsection{Capacity Window Effectiveness and Parameter Sensitivity}
\label{sec:capacity-window-effectiveness}

\subsubsection{Capacity Window Effectiveness}

\begin{figure}[t]
  \centering
  \begin{minipage}[b]{0.9\linewidth}
    \centering
    \includegraphics[width=\linewidth, trim=5 8 5 5, clip]{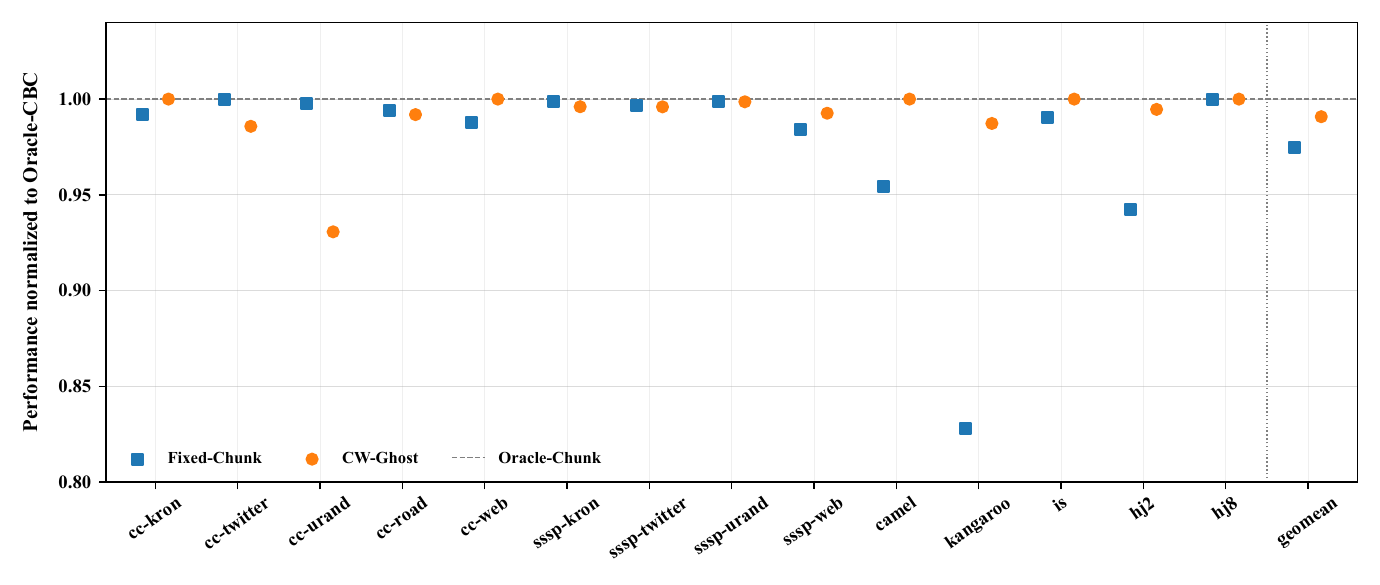}
    \textnormal{(a) Intel Xeon Gold 6258R}
  \end{minipage}

  \begin{minipage}[b]{0.9\linewidth}
    \centering
    \includegraphics[width=\linewidth, trim=5 8 5 5, clip]{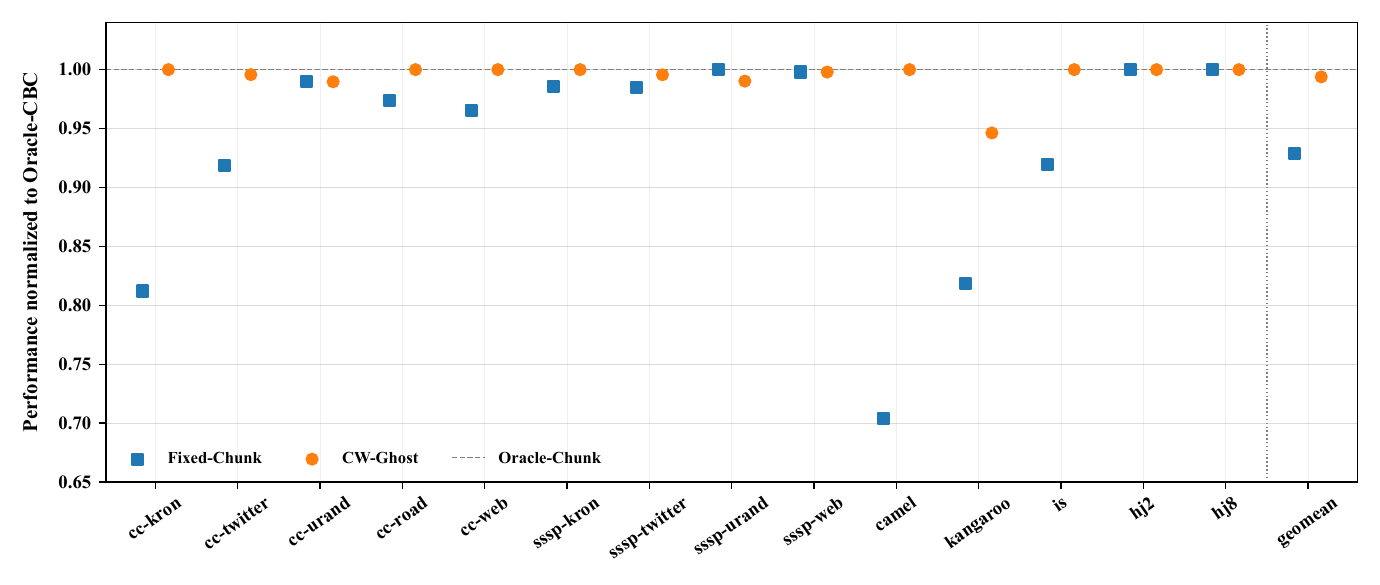}
    \textnormal{(b) AMD EPYC 7H12}
  \end{minipage}

  \caption{Performance of Fixed-Chunk and CW-Ghost normalized to
    Oracle-Chunk. Fixed-Chunk uses the best single granularity values of 64 and 32, respectively, shared across workloads on the Intel and AMD platforms.}
  \label{fig:capacity-window-effectiveness}
\end{figure}

Figure~\ref{fig:capacity-window-effectiveness} compares the performance
of Fixed-Chunk and CW-Ghost normalized to Oracle-Chunk. On the Intel and
AMD platforms, CW-Ghost achieves \(99.08\%\) and \(99.39\%\),
respectively, of Oracle-Chunk's geometric mean performance. In
comparison, the platform-tuned Fixed-Chunk configuration achieves
\(97.49\%\) and \(92.90\%\), respectively. CW-Ghost therefore improves
geometric mean performance over Fixed-Chunk by approximately \(1.6\%\)
on Intel and \(7.0\%\) on AMD. The platform-level fixed granularity remains competitive on Intel, but
incurs a more pronounced performance loss on AMD. In contrast, the
geometric mean performance gap between CW-Ghost and Oracle-Chunk is less
than \(1\%\) on both platforms. 

\subsubsection{Parameter Sensitivity}

\begin{figure}
  \centering
  \includegraphics[width=0.9\linewidth, trim=5 8 5 9, clip]
  {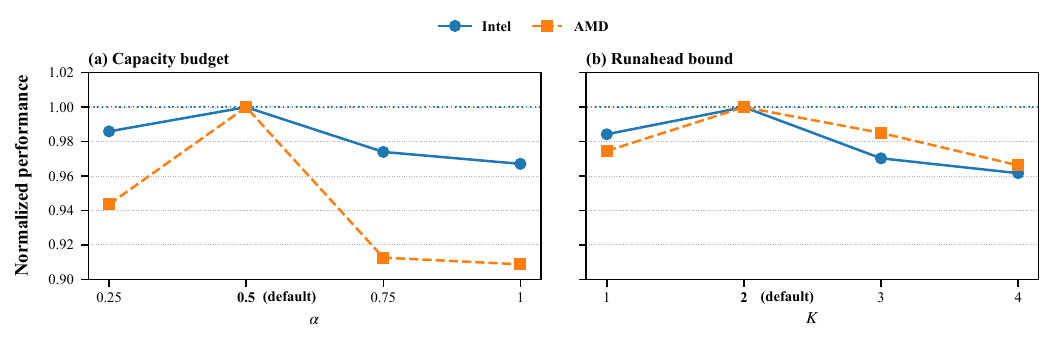}
  \caption{Sensitivity of CW-Ghost to (a) the capacity-budget ratio
    \(\alpha\) and (b) the CBC-level synchronization bound \(K\).
    Performance is normalized to the default configuration
    \((\alpha=0.5, K=2)\).}
  \label{fig:parameter-sensitivity}
\end{figure}

Figure~\ref{fig:parameter-sensitivity} evaluates the effects of the
capacity budget ratio \(\alpha\), and the CBC-level synchronization
bound \(K\), using five representative workloads:
\texttt{cc-twitter}, \texttt{sssp-web}, \texttt{camel}, \texttt{is},
and \texttt{hj2}. When varying one parameter, we hold the other at its
default value. The default configuration, \(\alpha=0.5\) and \(K=2\), achieves the
highest geometric mean performance among the evaluated settings on both
platforms. Relative to the default setting, changing \(\alpha\) reduces
geometric mean performance by \(1.4\%\)--\(3.3\%\) on Intel and by
\(5.6\%\)--\(9.1\%\) on AMD, indicating that the AMD platform is more
sensitive to the capacity budget ratio. By comparison, performance is
less sensitive to \(K\): among the non-default settings, the maximum
geometric-mean performance degradation is \(3.8\%\) on Intel and
\(3.4\%\) on AMD.

\subsection{Execution Overhead and Cache Behavior}
\label{sec:execution-overhead-cache-behavior}

Table~\ref{tab:normalized-cache-mpki} summarizes the cross-workload geometric means of the process-level retired instruction count and main-thread cache MPKI for Ghost Threading and CW-Ghost, normalized to Baseline on the same platform. MPKI denotes the number of cache-miss events per thousand retired instructions. Lower values are better.

\subsubsection{Dynamic Instruction Overhead}

On the Intel and AMD platforms, the geometric mean instruction counts of
CW-Ghost are \(1.10\times\) and \(1.42\times\), respectively, compared
with \(1.23\times\) and \(1.69\times\) for Ghost Threading. CW-Ghost therefore
reduces the retired instruction count relative to Ghost Threading by approximately
\(10.6\%\) on Intel and \(16.0\%\) on AMD. This reduction is consistent with the design goal of avoiding
redundant helper-thread execution.

\subsubsection{Cache Behavior}

For L1D misses,
the geometric-mean MPKI values of CW-Ghost are \(0.54\times\) and
\(0.74\times\) on the Intel and AMD platforms, respectively, compared
with \(0.63\times\) and \(0.86\times\) for Ghost Threading. Thus, CW-Ghost further
reduces L1D-miss MPKI relative to Ghost Threading by approximately \(14\%\) on
both platforms.
For LLC misses, the geometric-mean MPKI values of CW-Ghost are
\(0.34\times\) on Intel and \(0.31\times\) on AMD, compared with
\(0.40\times\) and \(0.38\times\) for Ghost Threading. These values correspond to
further reductions of approximately \(15\%\) and \(18\%\), respectively,
relative to Ghost Threading.

\begin{table}
    \centering
    \caption{Normalized retired instruction count and cache MPKI.}
    \label{tab:normalized-cache-mpki}
    \begin{tabular}{|l|c|c|c|c|}
        \hline
        \multirow{2}{*}{\textbf{Metric}} 
        & \multicolumn{2}{c|}{\textbf{Intel}}
        & \multicolumn{2}{c|}{\textbf{AMD}} \\
        \cline{2-5}
        & \textbf{Ghost Threading} & \textbf{CW-Ghost} & \textbf{Ghost Threading} & \textbf{CW-Ghost} \\
        \hline
        Retired instructions
        & \(1.23\times\)
        & \(1.10\times\)
        & \(1.69\times\)
        & \(1.42\times\) \\
        \hline
        L1D-miss MPKI
        & \(0.63\times\)
        & \(0.54\times\)
        & \(0.86\times\)
        & \(0.74\times\) \\
        \hline
        LLC-miss MPKI
        & \(0.40\times\)
        & \(0.34\times\)
        & \(0.38\times\)
        & \(0.31\times\) \\
        \hline
    \end{tabular}
\end{table}

Overall, CW-Ghost reduces the process-level dynamic instruction volume
and lowers the main thread's cache MPKI in terms of the geometric mean.
Both trends are consistent with the end-to-end performance results
reported in Section~\ref{sec:end-to-end-performance}.

\subsection{Configuration Cost, Applicability, and Limitations}
\label{sec:cost-applicability-limitations}

\subsubsection{Configuration Cost}
Oracle-Chunk enumerates 12 candidate CBC granularities and executes each
configuration three times to select the empirically optimal granularity
for each workload--platform pair, requiring 36 performance runs per
pair. In contrast, CW-Ghost requires only one offline profiling run per
workload--platform pair. It computes \(W_C\) from the sampling results
and uses it to configure the final optimized program, eliminating the need to test each candidate granularity individually.
CW-Ghost therefore substantially reduces the number of configuration
trials required to obtain a near-optimal CBC configuration.

\subsubsection{Applicability and Limitations}
CW-Ghost applies to programs with a clearly identified target
memory access region, future target addresses that can be computed ahead
of the main thread by a helper thread, and a target iteration space that
can be partitioned into CBCs. Its effectiveness depends on whether the
offline profile is representative of the memory access behavior during
deployment. Substantial changes in the input distribution, program
phase, or target processor require the workload to be reprofiled and
\(W_C\) to be recomputed.

Contention between the main and helper threads for SMT-shared resources
may offset part of the prefetching benefit. Moreover, the \(W_C\)
derived from the capacity model does not guarantee empirically optimal
performance for every workload. The current prototype automates data processing for the profile, \(W_C\) computation, and the parameterized
application of the CBC transformation; however, the target load and its
address computation slice must still be identified in advance.

\section{Conclusion}
\label{sec:conclusion}

Helper-thread prefetching requires not only an executable address computation slice for the target load but also an appropriate coverage of future iterations for each workload–platform pair. To address the limited adaptability of fixed CBC granularities, the unreliable transfer of granularity configurations across processors, and the cost of candidate-by-candidate search, we propose CW-Ghost. CW-Ghost derives a Capacity Window from a single offline profiling run, uses it to configure CBC granularity, and bounds helper-thread execution at the CBC level.

On both real system platforms, CW-Ghost improves the end-to-end performance of Ghost Threading, reduces the process-level retired instruction count, and lowers the main thread’s normalized cache MPKI. Without workload-specific search over candidate granularities, CW-Ghost achieves more than $99\%$ of Oracle-Chunk’s geometric mean performance. These results show that a capacity constraint derived from dynamic cache fill behavior can replace candidate-by-candidate granularity search with a single profiling run while achieving helper-thread prefetching performance close to the empirical optimum within the candidate set.

\subsubsection{Acknowledgements.} This work was supported by the National Natural Science Foundation of China (No. 62272474 and 61561146395).

%
%
\bibliographystyle{splncs04}
\bibliography{refs}

@inproceedings{guo2025ghost,
  title={Ghost Threading: Helper-Thread Prefetching for Real Systems},
  author={Guo, Yuxin and Bhosale, Akshay and Bora, Utpal and Chadwick, Alexandra W and Erd{\H{o}}s, M{\'a}rton and Gabrielli, Giacomo and Jones, Timothy M},
  booktitle={Proceedings of the 58th IEEE/ACM International Symposium on Microarchitecture},
  pages={899--914},
  year={2025}
}

@inproceedings{jamilan2022apt,
  title={Apt-get: Profile-guided timely software prefetching},
  author={Jamilan, Saba and Khan, Tanvir Ahmed and Ayers, Grant and Kasikci, Baris and Litz, Heiner},
  booktitle={Proceedings of the Seventeenth European Conference on Computer Systems},
  pages={747--764},
  year={2022}
}

@inproceedings{kim2004physical,
  title={Physical experimentation with prefetching helper threads on Intel's hyper-threaded processors},
  author={Kim, Dongkeun and Liao, SS-W and Wang, Perry H and Del Cuvillo, Juan and Tian, Xinmin and Zou, Xiang and Wang, Hong and Yeung, Donald and Girkar, Milind and Shen, John Paul},
  booktitle={International Symposium on Code Generation and Optimization, 2004. CGO 2004.},
  pages={27--38},
  year={2004},
  organization={IEEE}
}

@article{ainsworth2019software,
  title={Software prefetching for indirect memory accesses: A microarchitectural perspective},
  author={Ainsworth, Sam and Jones, Timothy M},
  journal={ACM Transactions on Computer Systems (TOCS)},
  volume={36},
  number={3},
  pages={1--34},
  year={2019},
  publisher={ACM New York, NY, USA}
}

@inproceedings{zhang2024rpg2,
  title={Rpg2: Robust profile-guided runtime prefetch generation},
  author={Zhang, Yuxuan and Sobotka, Nathan and Park, Soyoon and Jamilan, Saba and Khan, Tanvir Ahmed and Kasikci, Baris and Pokam, Gilles A and Litz, Heiner and Devietti, Joseph},
  booktitle={Proceedings of the 29th ACM International Conference on Architectural Support for Programming Languages and Operating Systems, Volume 2},
  pages={999--1013},
  year={2024}
}

@misc{beamer2015gap,
  author       = {Scott Beamer and Krste Asanović and David Patterson},
  title        = {The {GAP} Benchmark Suite},
  howpublished = {arXiv:1508.03619v3 [cs.DC]},
  year         = {2015},
  note         = {Version 3.},
  doi          = {10.48550/arXiv.1508.03619}
}

@inproceedings{luk2001tolerating,
  title={Tolerating memory latency through software-controlled pre-execution in simultaneous multithreading processors},
  author={Luk, Chi-Keung},
  booktitle={Proceedings of the 28th annual international symposium on Computer architecture},
  pages={40--51},
  year={2001}
}

@article{jung2006Helper,
  title={Helper thread prefetching for loosely-coupled multiprocessor systems},
  author={ Jung, Changhee  and  Solihin, Y.  and  Lim, Daeseob  and  Lee, Jaejin },
  journal={IEEE},
  year={2006},
}

@inproceedings{kamruzzaman2011inter,
  title={Inter-core prefetching for multicore processors using migrating helper threads},
  author={Kamruzzaman, Md and Swanson, Steven and Tullsen, Dean M},
  booktitle={Proceedings of the sixteenth international conference on Architectural support for programming languages and operating systems},
  pages={393--404},
  year={2011}
}

@manual{amd_uprof_2024,
  title        = {{uProf User Guide}},
  author       = {{Advanced Micro Devices, Inc.}},
  organization = {Advanced Micro Devices, Inc.},
  edition      = {Revision 4.2},
  year         = {2024},
  month        = jan,
  note         = {Publication \# 57368. \url{https://www.amd.com/en/developer/uprof/uprof-archives.html} (accessed 2024-10-01)}
}

@manual{intel_sdm,
  title        = {Intel{\textregistered} 64 and IA-32 Architectures Software Developer's Manual, Combined Volumes 1, 2A--2D, 3A--3D, and 4},
  author       = {{Intel Corporation}},
  organization = {Intel Corporation},
  year         = {2024}, 
  note         = {Order Number: 325462-084US, Revision 084. \url{https://www.intel.com/content/www/us/en/developer/articles/technical/intel-sdm.html} (accessed 2024-10-01)}
}

@article{bailey1991nas,
  author  = {Bailey, D. H. and Barszcz, E. and Barton, J. T. and others},
  title   = {The {NAS} Parallel Benchmarks},
  journal = {International Journal of Supercomputer Applications},
  volume  = {5},
  number  = {3},
  pages   = {63--73},
  year    = {1991}
}

@inproceedings{yu2015imp,
  title={IMP: Indirect memory prefetcher},
  author={Yu, Xiangyao and Hughes, Christopher J and Satish, Nadathur and Devadas, Srinivas},
  booktitle={Proceedings of the 48th International Symposium on Microarchitecture},
  pages={178--190},
  year={2015}
}

@inproceedings{navarro2022berti,
  title={Berti: an accurate local-delta data prefetcher},
  author={Navarro-Torres, Agust{\'\i}n and Panda, Biswabandan and Alastruey-Bened{\'e}, Jes{\'u}s and Ib{\'a}{\~n}ez, Pablo and Vi{\~n}als-Y{\'u}fera, V{\'\i}ctor and Ros, Alberto},
  booktitle={2022 55th IEEE/ACM International Symposium on Microarchitecture (MICRO)},
  pages={975--991},
  year={2022},
  organization={IEEE}
}

@inproceedings{hashemi2016continuous,
  title={Continuous runahead: Transparent hardware acceleration for memory intensive workloads},
  author={Hashemi, Milad and Mutlu, Onur and Patt, Yale N},
  booktitle={2016 49th Annual IEEE/ACM International Symposium on Microarchitecture (MICRO)},
  pages={1--12},
  year={2016},
  organization={IEEE}
}

@article{callahan1991software,
  title={Software prefetching},
  author={Callahan, David and Kennedy, Ken and Porterfield, Allan},
  journal={ACM SIGARCH Computer Architecture News},
  volume={19},
  number={2},
  pages={40--52},
  year={1991},
  publisher={ACM New York, NY, USA}
}

@article{mowry1992design,
  title={Design and evaluation of a compiler algorithm for prefetching},
  author={Mowry, Todd C and Lam, Monica S and Gupta, Anoop},
  journal={ACM Sigplan Notices},
  volume={27},
  number={9},
  pages={62--73},
  year={1992},
  publisher={ACM New York, NY, USA}
}

@inproceedings{fu2025magellan,
  title={Magellan: A High-Performance Loop-Guided Prefetcher for Indirect Memory Access},
  author={Fu, Gelin and Xia, Tian and Yin, Mingzhuo and Nair, Prashant J and Lis, Mieszko and Ren, Pengju},
  booktitle={Proceedings of the 52nd Annual International Symposium on Computer Architecture},
  pages={601--615},
  year={2025}
}

@inproceedings{naithani2020precise,
  title={Precise runahead execution},
  author={Naithani, Ajeya and Feliu, Josu{\'e} and Adileh, Almutaz and Eeckhout, Lieven},
  booktitle={2020 IEEE International Symposium on High Performance Computer Architecture (HPCA)},
  pages={397--410},
  year={2020},
  organization={IEEE}
}

@inproceedings{fu2024differential,
  title={Differential-matching prefetcher for indirect memory access},
  author={Fu, Gelin and Xia, Tian and Luo, Zhongpei and Chen, Ruiyang and Zhao, Wenzhe and Ren, Pengju},
  booktitle={2024 IEEE International Symposium on High-Performance Computer Architecture (HPCA)},
  pages={439--453},
  year={2024},
  organization={IEEE}
}

@inproceedings{luk1996compiler,
  title={Compiler-based prefetching for recursive data structures},
  author={Luk, Chi-Keung and Mowry, Todd C},
  booktitle={Proceedings of the seventh international conference on Architectural support for programming languages and operating systems},
  pages={222--233},
  year={1996}
}

@inproceedings{bera2021pythia,
  title={Pythia: A customizable hardware prefetching framework using online reinforcement learning},
  author={Bera, Rahul and Kanellopoulos, Konstantinos and Nori, Anant and Shahroodi, Taha and Subramoney, Sreenivas and Mutlu, Onur},
  booktitle={MICRO-54: 54th Annual IEEE/ACM International Symposium on Microarchitecture},
  pages={1121--1137},
  year={2021}
}

@inproceedings{ainsworth2024triangel,
  title={Triangel: A high-performance, accurate, timely on-chip temporal prefetcher},
  author={Ainsworth, Sam and Mukhanov, Lev},
  booktitle={2024 ACM/IEEE 51st Annual International Symposium on Computer Architecture (ISCA)},
  pages={1202--1216},
  year={2024},
  organization={IEEE}
}

@article{Guo2018DSAP,
  author  = {Guo, Hui and Huang, Libo and L{\"u}, Ya-Shuai
             and Ma, Jianqiao and Qian, Cheng and Ma, Sheng
             and Wang, Zhiying},
  title   = {Accelerating {BFS} via Data Structure-Aware
             Prefetching on {GPU}},
  journal = {IEEE Access},
  volume  = {6},
  pages   = {60234--60248},
  year    = {2018},
  doi     = {10.1109/ACCESS.2018.2876201}
}

@article{Lu2021GraphPEG,
  author  = {L{\"u}, Ya-Shuai and Guo, Hui and Huang, Libo
             and Yu, Qi and Shen, Li and Xiao, Nong
             and Wang, Zhiying},
  title   = {{GraphPEG}: Accelerating Graph Processing on {GPUs}},
  journal = {ACM Transactions on Architecture and Code Optimization},
  volume  = {18},
  number  = {3},
  pages   = {1--24},
  articleno = {30},
  year    = {2021},
  doi     = {10.1145/3450440}
}

\end{document}